\documentclass[preprint,12pt]{elsarticle}

\def\svn$#1: #2 #3 #4 #5 #6${\def\revision{Revision #3, #4}}
\svn $Id: pottsgpgpu.tex 2396 2010-12-31 04:17:26Z ferrero $

\usepackage{graphicx}
\usepackage{epsfig}
\usepackage{amssymb}
\usepackage{amsthm}

\usepackage[english]{babel}
\usepackage[utf8]{inputenc}
\usepackage[usenames]{color}
\usepackage{url}
\usepackage{hyperref}
\usepackage[all]{hypcap}
\usepackage[normalem]{ulem}

\newcommand{\code}[1]{\texttt{#1}}

\journal{Computer Physics Communications}

\begin{document}

\begin{frontmatter}

\title{q-state Potts model metastability study using optimized GPU-based Monte Carlo algorithms}

\author[1]{Ezequiel E. Ferrero}
\author[2]{Juan Pablo De Francesco}
\author[2]{Nicol\'as Wolovick}
\author[2,3]{Sergio A. Cannas}

\address[1]{CONICET, Centro Atómico Bariloche, 8400 San Carlos de Bariloche, R\'io Negro, Argentina}

\address[2]{Facultad de Matem\'atica, Astronom\'{\i}a y
F\'{\i}sica, Universidad Nacional de C\'ordoba,
Ciudad Universitaria, 5000 C\'ordoba, Argentina}

\address[3]{Instituto de F\'{\i}sica Enrique Gaviola (IFEG-CONICET),
Ciudad Universitaria, 5000 C\'ordoba, Argentina}

\begin{abstract}
We implemented a GPU based parallel code to perform Monte Carlo simulations of the two dimensional q-state Potts model. The algorithm is based on a checkerboard update scheme and assigns independent random numbers generators to each thread. The implementation allows to simulate systems up to $\sim 10^9$ spins with an average time per spin flip of $0.147ns$ on the fastest GPU card tested, representing a speedup up to 155x, compared with an optimized serial code running on a high-end CPU.

The possibility of performing high speed simulations at large enough system sizes  allowed us to provide a positive numerical evidence about the existence of metastability on very large systems based on Binder's criterion, namely, on the existence or not of specific heat singularities at spinodal temperatures different of the transition one.
\end{abstract}


\end{frontmatter}

\section{Introduction}
\label{Intro}

The tremendous advances allowed by the usage of numerical simulations in the last decades have promoted these techniques to the status of  indispensable tools in modern Statistical Mechanics research. Notwithstanding, many important theoretical problems in the field still remain difficult to handle due to limitations in the available computational capabilities. Among many others, typical issues that challenge the numerical treatment concern systems with slow dynamics (i.e., dynamical processes that involve very different time scales) and/or strong finite size effect, which require fast simulations of a very large number of particles. Some typical examples we may cite are spin glass transitions~\cite{Fisher-Hertz-book1993}, glassy behavior~\cite{Ko2003,Binder-Kob-book2005} and grain growth~\cite{Cu2010}. In such kind of problems the state of the art is usually launched by novel numerical approaches or extensive computer simulations. In this sense, the advent of massive parallel computing continuously opens new possibilities but, at the same time, creates a demand for new improved algorithms. In particular, the usage of GPU cards (short for Graphics Processing Units) as parallel processing
devices is emerging as a powerful tool for numerical simulations in Statistical Mechanics
systems~\cite{PrViPaSc2009,BlViPr2010,BePaPa2011,HaLePl2010,We2010,We2011}, as well as in other areas of
physics~\cite{HeSiBeGuTi2010,Ti2010,ClBaBaBrRe2010}.

These GPUs have a Toolkit that abstracts the end-user from many low-level implementation details, yet all the typical problems of concurrency exists and they are magnified by the massive amount of (virtual) threads it is capable to handle.
An extremely fine grained concurrency is possible and advised thanks to the Single Instruction Multiple Thread (SIMT) model. Therefore, any non trivially independent problem requires a correct concurrency control (synchronization), and the lack of it hinders correctness in a much dramatic way than current 4 or 8-way multicore CPU systems.
The other challenge apart from correctness is performance, and here is where the algorithm design practice excels.
Taking into account internal memory structure, memory/computation ratio, thread division into blocks and thread internal state size, can boost the algorithm performance ten times from a trivial implementation~\cite{Ryoo08}.
It is also customary to give an approximation of the speedup obtained from a CPU to GPU implementation in terms of ``$N$x'',
 even though, as we discuss later, this number will always depend on the corresponding efforts devoted to optimally programming
for each architecture.

\medskip

In this work we focus on GPU based Statistical Mechanics simulations of lattice spin systems. In particular, we study the metastability problem in the ferromagnetic $q$-state Potts model~\cite{Wu1982} in two dimensions when $q > 4$.
While  this phenomenon is clearly observed in finite size systems, its persistence in the thermodynamics limit is still an  unsolved problem and subject of debate~\cite{Bi1981,MeMo2000,PeIbLo2008,BaBeDu2008,LoFeGrCa2009}.
In an earlier work, Binder proposed a numerical criterion to determine whether metastability remains in the thermodynamic limit or not, based on the  scaling properties of the average energy in the vicinity of the transition temperature~\cite{Bi1981}.
However, the narrow range of temperature values of the metastable region requires high precision calculations for the criterion to work.
Hence, to reduce finite size bias and  statistical errors down to an appropriated level, large enough system sizes are needed. The computation capabilities required to carry out such calculations in a reasonable time were unavailable until recently.

We developed an optimized algorithm to perform Monte Carlo numerical simulations of the $q$-state Potts model on GPU cards.
This algorithm allowed us to simulate systems up to $N=32768\times32768 \sim 1.073 \times 10^9$ spins with a lower bound time of 0.147ns per spin flip using using an NVIDIA GTX 480 Fermi card,
and in terms of speedup, we obtained 155x from an optimized CPU sequential version running on an Intel Core 2 Duo E8400 at 3.0GHz.

What is remarkable about the speedup is that it allowed us to explore bigger systems, simulate more iterations, explore parameters in a finer way, and all of it at a
relatively small
cost in terms of time, hardware and coding effort.
With this extremely well performing algorithm we obtained a positive numerical evidence of the persistence of metastability in the thermodynamic limit for $q>4$, according to Binder's criterion.

\smallskip

The paper is structured as follows.
In Section~\ref{model} we briefly review the main properties of the Potts model and the particular physical problem we are interested in. In Section~\ref{algorithm} we introduce the simulation algorithm and in Section~\ref{validate} we compare the predictions of our numerical simulations against some known equilibrium properties of the model to validate the code.
In Section~\ref{performance} we check the performance of the code. In Section~\ref{metastability} we present our numerical results concerning the metastability problem.
Some discussions and conclusions are presented in Section~\ref{discussion}.

\section{The q-state Potts model}
\label{model}

\subsection{The model}
The $q$-state Potts model~\cite{Wu1982} without external fields is defined by the Hamiltonian

\begin{equation}
H = - J \sum_{<i,j>} \delta (s_i,s_j) \label{Hamiltonian}
\end{equation}

\noindent where $s_i=1,2,\ldots,q$, $\delta (s_i,s_j)$ is the
Kronecker delta and the sum runs over all nearest neighbors pairs of spins in a Bravais lattice with $N$ sites. Being a generalization of the Ising model ($q=2$), this model displays a richer behavior than the former. One of the main interests  is that the two-dimensional ferromagnetic version ($J>0$) exhibit a first order phase transition at some finite temperature when $q>4$, while for $q\leq 4$ the transition is continuous~\cite{Wu1982}. Hence, it has become a paradigmatic model in the study of phase
transitions and their associated dynamics, like for instance,  domain growth kinetics~\cite{ViGr1987,GrAnSr1988,SiMa1995b,FeCa2007,LoArCuSi2010} and
nucleation as an equilibration mechanism~\cite{MeMo2000, Ru2002, BaGuTrTrHu2010}.

Some equilibrium properties of the two-dimensional model are known exactly, which allows  numerical algorithms testing. We list here some of them that are used for comparison with the numerical results in the present work. For instance, the transition temperature for the square lattice in the thermodynamic limit is given by~\cite{Ba1973}

\begin{equation}\label{Tc}
    \frac{k_B T_c}{J} = \frac{1}{\ln (1+\sqrt{q})}
\end{equation}

\noindent where $k_B$ is the Boltzmann constant. Hereafter we will choose $k_B/J=1$.
Considering the energy per spin $e=\langle H \rangle /N$, in the thermodynamic limit
the latent heat for $q>4$ is~\cite{Ba1973}

\begin{equation}\label{Ujump}
    e_d - e_o = 2 \left(1+\frac{1}{\sqrt{2}} \right)\, \tanh\frac{\Theta}{2}\,\prod_{n=1}^\infty (\tanh\, n\Theta)^2
\end{equation}

\noindent where $\Theta=\arccos\, \sqrt{q}/2$ and

\begin{eqnarray}\label{Ucpm}
    e_d= \lim_{N\to\infty} \frac{1}{N} \lim_{T\to T_c^+} \langle H \rangle,\\
    e_o= \lim_{N\to\infty} \frac{1}{N} \lim_{T\to T_c^-} \langle H \rangle.
\end{eqnarray}

\noindent Also

\begin{equation}\label{Uadd}
    e_d + e_o = -2(1+1/\sqrt{q})
\end{equation}

\noindent from which the individual values of $e_d$ and $e_o$ can be obtained~\cite{KiMiSh1954}.

The order parameter is defined as
\begin{equation}\label{magnetization}
 m = \frac{q(N_{max}/N - 1)}{q-1}
\end{equation}
where $N_{max} =max(N_1, N_2,\ldots,N_q)$, being $N_i$ the number of spins in state $i$. At the transition the jump in the order parameter (for $q>4$)  is given by~\cite{Ba1982}

\begin{equation}
 \Delta m = 1- q^{-1} - 3q^{-2} - 9q{-3} - 27q^{-4} - \ldots
 \label{magne-exact}
\end{equation}

\subsection{Metastability}

The problem of metastability in the infinite size $q$-state Potts model (for $q>4$) is an old standing problem in statistical mechanics ~\cite{Bi1981,VeBeHe2003,FeCa2007,IbLoPe2007,PeIbLo2008,LoFeGrCa2009}. It has also
kept the attention of the Quantum Chromodynamics' (QCD) community for many years~\cite{Me1996, KaSt2000, VeBeHe2003,BaBeDu2008,BoEl2010}, because it has some characteristics in common with the deconfining (temperature driven) phase transition in heavy quarks.

Metastability is a verified fact in a finite system. It is known~\cite{MeMo2000,FeCa2007,PeIbLo2008} that below but close to $T_c$ the system quickly relaxes to a disordered (paramagnetic) metastable state, with a life time that diverges as the quench temperature $T$ approaches $T_c$ (see for example Fig.4 in Ref.\cite{LoFeGrCa2009}). This state is indistinguishable from one in equilibrium in the sense of local dynamics, namely, two times correlations depends only on the difference of times, while one time averages are stationary~\cite{PeIbLo2008}.

Nevertheless, the existence of metastability in the thermodynamic limit is still an open problem~\cite{PeIbLo2008}.
In Ref.\cite{Bi1981} Binder studied static and dynamic critical behavior of the model
(\ref{Hamiltonian}) for $q=3,4,5,6$. Using standard Monte Carlo procedures he obtained
good  agreement with exact results for energy and free energy at the critical point and critical exponents estimates for $q=3$ in agreement with high-temperature series extrapolations and real space renormalization-group methods.
When analyzing the $q=5$ and $6$ cases he realized that the transition is, in fact, a very
weak first order transition, where pronounced ``pseudocritical'' phenomena occur.
He studied system sizes from $N=16\times16$ up to $N=200\times200$, and observation times
up to $10^4 \mathit{MCS}$ (a Monte Carlo Step $\mathit{MCS}$ is defined as as a complete cycle of $N$ spin update trials, according
to the Metropolis  algorithm). Within his analysis he was unable to distinguish between two different
scenarios for the transition at $q \geq 5$ due to finite size effects taking place at
the simulations. He proposed two self-avoiding possible scenarios for the transition.
 In the first one the energy per
spin reaches the transition temperature with a finite slope both coming from higher and
lower temperatures, thus projecting metastable branches at both sides of the transition that
end at temperatures $T_{sp}^+$ and $T_{sp}^-$ both different from $T_c$.
In the second scenario, the energy reaches $T_c$
with an infinite slope which would imply a first order phase transition with a true divergence
of the specific heat at $T_c$.

On the other hand, other approaches based on different definitions of the spinodal temperatures predict, either the convergence of the finite size spinodal temperatures to $T_c$  \cite{MeMo2000,BaBeDu2008} or a convergence to limit values different from but closely located to $T_c$ \cite{LoFeGrCa2009}.

\section{Optimized GPU-based Monte Carlo algorithm for the q-state Potts model}
\label{algorithm}

We developed a GPU based code to simulate the 
two dimensional Potts model, using classical Metropolis dynamics
 on square lattices of size $N= L \times L$ sites
with periodic boundary conditions. For the spin update we  partition lattice sites in two sets, the whites and the blacks, laid out in a framed checkerboard pattern in order to update in a completely asynchronous way all the white cells first and then all the black ones
(given that the interactions are confined to nearest neighbors).
This technique is also know as the Red-Black Gauss-Seidel~\cite{pres92nrinc}.
We analyzed equilibrium states of systems ranging from
$N=16\times16$ to $N=32768\times 32768$ ($2^{15}\times2^{15} \simeq 1.073\times 10^9$ spins).

The typical simulation protocol is the following. Starting from an initial
ordered state ($s_i=1$ $\forall i$) we fix the temperature to $T=T_{min}$ and run
$t_{\mathit{tran}}$ to attain equilibrium, then we run
$t_{\mathit{max}}$ taking one measure each $\delta t$ steps
to perform averages.
After that, we keep the last configuration of the system
and use it as the initial state for the next temperature, $T=T_{\mathit{min}}+\delta T$. This process is repeated until some maximum temperature  $T_{\mathit{max}}$ is reached.
We repeat the whole loop for several samples to average over different realizations
of the thermal noise.
In a similar way we perform equilibrium measurements going from $T_{\mathit{max}}$ to $T_{\mathit{min}}$ starting initially from a completely random state.

\subsection{GPU: device architecture and CUDA programming generalities}

In 2006, NVIDIA decided to take a new route in GPU design and launched the G80 graphics processing unit, deviating from the standard pipeline design of previous generations and transforming the GPU in an almost general purpose computing unit.
Although this decision could have been driven by the gaming community asking for more frames per second, NVIDIA took advantage of his General Purpose Graphics Processing Units (GPGPU), and in 2007 they launched the CUDA SDK, a software development kit tailored to program its G80 using C language plus minor extensions.
The G80 hardware and the CUDA compiler quickly proved to have an extremely good relation in terms of GFLOPS per watt and GFLOPS per dollar with respect to the CPU alternatives in the application field of numerical algorithms.

The architecture has evolved two generations, GT200 in 2008 and 2009, and the GF100 in 2010,
also known as the Fermi architecture.
All of them share the same Single Instruction Multiple Thread (SIMT) concurrency paradigm in order to exploit the high parallelism (up to 480 computing cores) and the high memory bandwidth (up to 177GBps).
The SIMT model is a convenient abstraction that lies in the middle of the SIMD (Single Instruction Multiple Data) and MIMD (Multiple Instruction Multiple Data), where the first reigned in the 80's with the vector computers, and the later is the commonplace of almost every computing device nowadays, from cellphones to supercomputers.

Using SIMT paradigm, the parallel algorithm development changes greatly since it is possible to code in a one-thread-per-cell fashion.
The thread creation, switching and destruction have such a low performance impact that doing a matrix scaling reduces to launch one kernel per matrix cell, even if the matrix is $32768\times 32768$ of single precision floating point numbers summing up 1 GThread all proceeding in parallel.
In fact, for the implementation, the more threads the better, since the high memory latency to global memory (in the order of 200 cycles) is hidden by swapping out warps (vectors of 32 threads that execute synchronously) waiting for the memory to become available.

It is important to emphasize the role of blocks in the SIMT model.
Threads are divided into blocks, where each block of threads have two special features: a private shared memory and the ability to barrier synchronize.
Using these capabilities, the shared memory can be used as a manually-managed cache that in many cases greatly improves the performance.

We used the GTX 280, GTX 470 and GTX 480 boards. The relevant hardware parameters for these boards are shown in table~\ref{hardware-parameters-table}.

\begin{table}[h!]
\label{hardware-parameters-table}
\begin{small}
\begin{center}
\begin{tabular}{|l|r|r|r|}
	\hline
	Board Model & GTX 280 & GTX 470 & GTX 480\\
	\hline
	Available & Q2 2008 & \multicolumn{2}{c|}{Q1 2010} \\
	GPU & GT200 & \multicolumn{2}{c|}{GF100} \\
	CUDA capability & 1.3 & \multicolumn{2}{c|}{2.0} \\
	\hline
	CUDA cores & 240 & 448 & 480 \\
	Processor Clock & 1.30GHz & 1.22GHz & 1.40GHz \\
	Global Memory & 1GB & 1.25GB & 1.50GB \\
	Memory Bandwidth & 141.7GBps & 133.9GBps & 177.4GBps \\
	\hline
	L1 Cache & N/A & \multicolumn{2}{c|}{16KB-48KB}  \\
	L2 Cache & N/A & \multicolumn{2}{c|}{768KB} \\
	\hline
	Max \# of Threads per Block & 512 & \multicolumn{2}{c|}{1024} \\
	Shared Memory per Block & 16KB & \multicolumn{2}{c|}{48KB-16KB} \\
	Max \# of Registers per Block & 16384 & \multicolumn{2}{c|}{32768} \\	
	\hline
\end{tabular}
\caption{Key features about NVIDIA GTX 280, GTX 470, and GTX 480 graphic cards.}
\end{center}
\end{small}
\end{table}

The improvements of the Fermi architecture lies on the new computing capabilities (improved Instruction Set Architecture -- ISA), the doubling of cores, the inclusion of L1 and L2 cache, increased per-block amount of parallelism and shared memory.


As every modern computing architecture the memory wall effect has to be relieved with a hierarchy of memories that become faster, more expensive and smaller at the top.
The bottom level is the global memory, accessible by every core and having from 1GB to 1.5GB of size\footnote{This values apply to consumer graphics cards. The Tesla HPC line incorporates up to 6GB of memory (e.g. Tesla C2070), that is configurable to be ECC in order to improve reliability} 
and a latency of 200 cycles.
The next level is the shared memory, that is configurable 16KB or 48KB per block having a latency of only 2 cycles.
At the top there are 32K registers per block.
There are also texture and constant memory, having special addressing capabilities, but they do not bring any performance improvement in our application.
 The Fermi architecture has also incorporated ECC memory support to eventually deal with internal data corruption.

The programming side of this architecture is a ``C for CUDA'', an extension of the C Programming Language~\cite{KR78} that enables the \emph{host} processor to launch \emph{device} kernels~\cite{Kirk10}.
A kernel is a (usually small) piece of code that is compiled by \texttt{nvcc}, the NVIDIA CUDA Compiler, to the PTX assembler that the architecture is able to execute.
The kernel is executed simultaneously by many threads, organized in a two-level hierarchic set of parallel instances indexed as $(\mathit{grid},\mathit{block})$ (a grid of thread blocks).
Internally each grid and block can be divided up to two dimensions for the grid and three dimensions for the block, in order to establish a simple thread-to-data mapping.
Special variables store the thread position information of block and thread identifier $(\mathit{bid},\mathit{tid})$ that distinguishes the threads executing the kernel.

It is interesting to note that although the unit of synchronous execution is a warp of 32 threads, the threads inside a warp may diverge in their execution paths (occurrence of
bifurcations), at the cost of having to re-execute the warp once for each choice taken.
Needless to say that in general this impacts negatively in the performance and has to
be avoided.


The present code is divided in two main functions: spin update and energy and magnetization computation.
The first function is implemented in host code by the function \code{update} and this comprises calling the device kernel \code{updateCUDA} once updating white cells and next updating black cells in a checkerboard scheme.
The energy and magnetization (and their related moments) summarization is done by \code{calculate} that calls the kernel \code{calculateCUDA} and two more auxiliary kernels: \code{sumupECUDA} and \code{sumupMCUDA}.

\subsection{Random Number Generator}

The Potts model simulation requires a great amount of random numbers. Namely, each cell
updating its spin needs one integer random number in $\{0,\dots, q\!-\!1\}$ and possibly  a second one in the real range $[0,1)$ to decide the acceptance of the flip.
Hence, a key factor to performance is using a good parallel random number generator.

Given the great dependence in terms of time (it has to be fast) and space (small number of per-thread variables), we find Multiply-With-Carry (MWC)~\cite{Marsaglia:2003:RNG} ideal in both aspects.
Its state is only $64$ bits, and obtaining a new number amounts to compute $x_{n+1} = (x_n \times a + c_n)\bmod{b}$, where $a$ is the multiplier, $b$ is the base, and $c_n$ is the carry from previous modulus operation.
We took the implementation from the CUDAMCML package~\cite{cudamcml_rng} that fixes $b=2^{32}$ in order to use bit masks for modulus computation.

For independent random number sequences, MWC uses different multipliers, and they have to be \emph{good} in the following sense: $a\times b-1$ should be a safeprime, where $p$ is a safeprime if both $p$ and $(p-1)/2$ are primes.
Having fixed $b=2^{32}$, the process of obtaining safe primes boils down to test for primality of two numbers $\mathit{goodmult}(a) \equiv \mathit{prime}(a\times 2^{32}-1) \land \mathit{prime}((a\times 2^{32}-2)/2)$.
It is important to remark that the nearer to $2^{32}$ is $a$ the longer the period of the MWC (for $a$ close to its maximum, the period is near to $2^{64}$), therefore it is always advisable to start looking for $\mathit{goodmult}$ down from $2^{32}-1$.

We limit the amount of independent random number generators (RNG) to $512^2/2 = 131072$ that is slightly lower than the $150000$ good multipliers that CUDAMCML gives in its file \code{safe\_primes\_base32.txt}.
The state needed comprises 12 bytes per independent RNG, totalizing 1.5MB of global memory, less that $0.15\%$ of the total available in the GTX 280.
We consider this a good trade-off between independence in number generation and memory consumption.
This design decision is crucial in the parallelization of the spin update function, as we frame the lattice in rectangles of $512\times 512$, to give each thread an independent
RNG\footnote{For system sizes smaller than $N=512^2$ we use smaller frames,
and then, fewer RNG. But $512\times 512$ is the standard framing choice for most of the work.}.
Moreover, this implies that the larger the lattice, the more work will be done by a single thread.

It is important to remark we are well below the RNG cycle even for the largest simulations.

\subsection{Spin update}

\begin{figure}
\begin{center}
\includegraphics[scale=1]{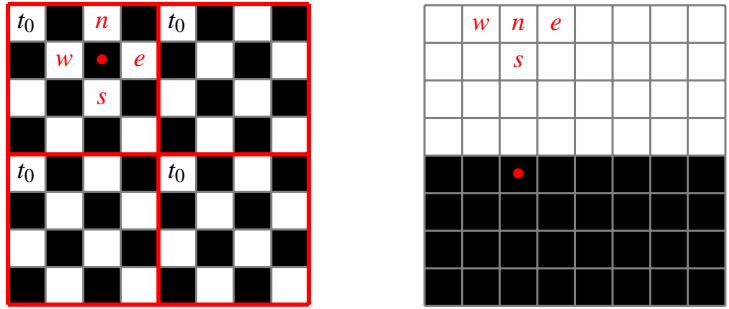}
\caption{On the left: a $8\times 8$ checkerboard framed in $4\times 4$ (red marking), the cells updated by thread $t_0$ are singled out, we also marked the north, east, south and west neighbors of cell $\bullet$.
	On the right: packed checkerboard showing first half of whites, where the neighboring cells $n,e,s,w$ are marked, also in the second half of black cells $\bullet$ is singled out.
}
\end{center}
\label{fig:checkerboard}
\end{figure}

On top of the checkerboard division we have first to frame the lattice in rectangles of $512\times 512$ in order to use the limited amount of independent RNG (Fig.\ref{fig:checkerboard}, left). This implies launching two consecutive kernels (black/white) of $512\times 512/2$ threads, typically organized into a grid of $32\times 16$ blocks of  $16\times 16$ threads.
The second step comprises the remapping of a two dimensional stencil of four points
in order to save memory transfers. The row-column pair $(i,j)$ is mapped to $(((i\!+\!j)\textit{mod}\ 2 \times L + i)/2, j)$, and this allows to pack all white and all black cells in contiguous memory locations improving locality and allowing wider reads of 3 consecutive bytes (Fig.\ref{fig:checkerboard}, right).

We encode each spin in a byte, allowing simulations with $q\leq256$ and $L^2\leq\mathit{available\ RAM}$.
Since some extra space is needed for the RNG state and for energy and magnetization summarization, this upper bound is not reached.
The biggest simulations we achieve is $L=32768, q=45$ for the GTX 480.

It is important to remark that shared memory is not used,
since 
we could not improve performance and it hindered readability of the code.
Texture memory techniques were not used for the same reasons.

\subsection{Computation of Energy and Magnetization}

During the evolution of the system we extract periodically two quantities:
energy Eq.(\ref{Hamiltonian}) and magnetization Eq.(\ref{magnetization}).
The kernel responsible for this job is \code{calculateCUDA}.
It first partitions the cells into CUDA blocks. In each block we have easy
access to barrier synchronization and shared memory among its threads.
Each block within its cells adds the local energies and accumulates in a partial
vector $(n_1, n_2,\ldots, n_q)$ the number of spins in each state. This is performed
in shared memory using atomic increments to avoid race conditions.
After that, those blocks' results are added up in parallel using a butterfly-like algorithm~\cite{Kirk10}
by kernels \code{sumupECUDA} and \code{sumupMCUDA}, but none of the known optimizations~\cite{harris07cuda} are applied, since it implies obfuscating the code for a marginal global speedup.
Previous kernels end up with up to approximately a thousand partial energies and vectors of spin counters, that are finally added in the CPU.

It has to be noticed that device memory consumption in this part is linear not only in $N$, but also in $q$.

\subsection{Defensive Programming Techniques and Code Availability}

Writing scientific code that is maintainable, robust and repeatable is of utmost importance for the fields of science
where computer simulation and experimentation is an everyday practice~\cite{merali10error}.

CUDA coding in particular is hard, not only in creating the algorithms, choosing a good block division and
trying to take advantage of all its capabilities, but also, in the debugging and maintenance cycle.
Debugging tools are evolving rapidly, for example there is a memory debugger \code{cuda-memcheck} that is shipped with current CUDA SDK.
Nevertheless, we would rather adhere to some passive and active security measures within our code
to make it easier to understand and modify, and at the same time, to make it robust in the sense of no
unexpected hangs, miscalculations or silent fails.

Among passive security measures, we use assertions (boolean predicates) related to hardware limitations like the
maximum of 512 threads per block.
Other use of the assertions is checking for the integer representation limitations: given the computing power that
GPGPU brings, lattices of $32768\times 32768$ are feasible to simulate, and integer overflow could be a possibility,
for example when computing the total energy.
Assertions were also used to enforce preconditions on algorithm running, for example, the spin updating cannot do
well if $L$ is not multiple of the frame size.
We also check every return condition of CUDA library calls and kernels, in order to lessen the asynchrony of error detection in CUDA.
The same practice is used in standard library calls for file handling.

Active security measures are also taken.
We use tight types in order to detect problems in compile time.
We also decorate parameters and variable names with \code{const} modifiers where applicable.
For pointer immutable parameters we forbid the modification of pointed data as well as the pointer itself.
The scope of automatic variables is as narrow as possible, declaring them within blocks,
in order to decrease the namespace size in every line of code.
We put in practice the simple but effective idea of using meaningful variable names in order to improve the readability.

We also adhere to the practice of publishing the code~\cite{barnes10publish} in the line of~\cite{PrViPaSc2009,BlViPr2010,We2010}, since it benefits from community debugging and development.
It can be found on~\cite{potts3site}.

\section{Algorithm checking}
\label{validate}

In order to validate our CUDA code we run some typical simulations to measure well established results.

\begin{figure}
\centering
\includegraphics[scale=0.45]{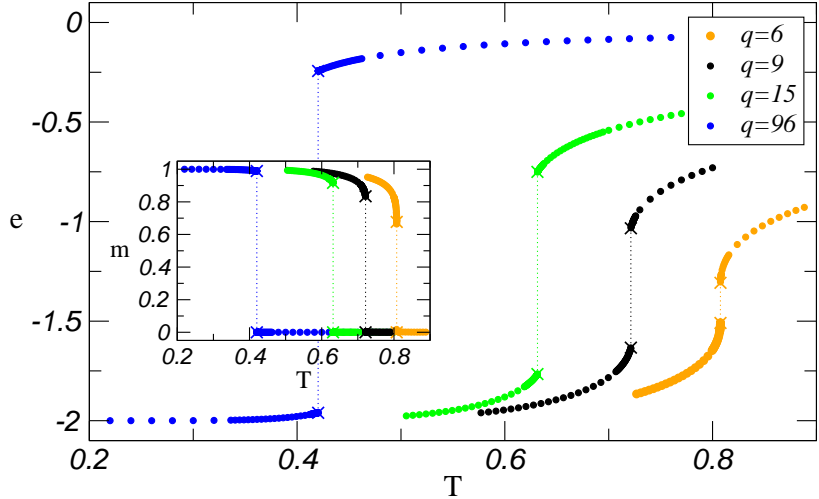}
\caption{(Color online) Equilibrium energy per spin $e$ and magnetization $m$ (inset) versus temperature for $q=9,12,15, 96$. Exact values at the transition point  from equations~(\ref{Ujump}), (\ref{Uadd}) and (\ref{magne-exact}) are marked as crosses. Data comes from averages over $10$ samples of linear system size $L=2048$. Error bars are smaller than the symbol size.}
\label{fig3}
\end{figure}

First we calculate the energy per spin $e$ and magnetization $m$ above and below the transition temperature, by cooling (heating) from an initially disordered (ordered) state. The behaviors of $e$ and $m$ as functions of $T$ for different values of $q$ are shown in Fig.\ref{fig3}. From these calculations we obtain the values of the energy ($e_d$ and $e_o$) and magnetization jump $\Delta m$ at the exact transition temperature (see Section~\ref{model}).
Results are compared with exact values in table~\ref{valores-medidos-vs-exactos}

\begin{table}\label{valores-medidos-vs-exactos}
\begin{small}
\begin{center}
\begin{tabular}{|c||c|c|c|c|c|c|}
	\hline
	$q$ & \multicolumn{2}{c|}{$- e_o$} & \multicolumn{2}{c|}{$- e_d$} & \multicolumn{2}{c|}{$\Delta m$}\\
	\cline{2-7}
	    & exact & calculated & exact & calculated & exact & calculated \\
	\hline\hline
	 6& 1.508980...& 1.51(2)   & 1.307516...& 1.306(1)  & 0.677083...& 0.674(2)\\
	\hline
	 9& 1.633167...& 1.6332(5) & 1.033499...& 1.0334(5) & 0.834019...& 0.8338(4)\\
	\hline
	15& 1.765905...& 1.7659(2) & 0.750492...& 0.7509(4) & 0.916693...& 0.9167(3)\\
	\hline
	96& 1.960306...& 1.96030(3)& 0.243817...& 0.24382(4)& 0.989247...& 0.98924(2)\\
	\hline
\end{tabular}
\caption{Comparison between calculated and known exact values of $e_o$, $e_d$, and $\Delta m$ at the transition for different values of $q$. Results were obtained from averages over $10$ samples of linear system size $L=2048$ and equilibration and measurements times of at least $5\times10^5$ MCS each
one.}
\end{center}
\end{small}
\end{table}

We can see a very good agreement between data and exact results.
It's worth noting that the data from table~\ref{valores-medidos-vs-exactos} is not the result of extrapolations of
some finite size analysis, but the values from curves in Fig.\ref{fig3} at the transition itself.
 Since we measure one point each $\Delta T$ in temperature, cooling and heating procedures won't necessary lead to a
point measured exactly at $T_c$. So, we have to interpolate points close to $T_c$ to deduce the corresponding values of
$e_o$, $e_d$ and $m_o$ at $T_c$. The differences obtained from interpolations using points separated by $\Delta T$
and points separated by $2\times \Delta T$ determine the estimated errors.

We also calculate the fourth order cumulant of the energy~\cite{Ja1993,Bi1997}

\begin{equation}
 V_L = 1- \frac{\langle H^4 \rangle}{3 \langle H^2 \rangle^2}
\end{equation}

\noindent as a function of the temperature for $q=6$ and different system sizes. As it is well known, $V_L$ is almost constant far away from
the transition temperature and exhibits a minimum at a pseudo critical temperature
\begin{equation}
T_c^*(L)  = T_c + \frac{T_c^2 \ln(q e_o^2 / e_d^2)}{e_d - e_o} \frac{1}{L^d}
\end{equation}

%

\begin{figure}
\centering
\includegraphics[scale=0.45]{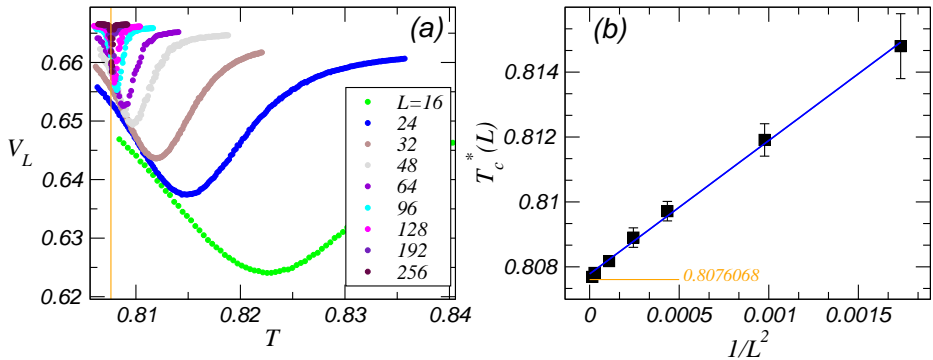}
\caption{ (Color online) Finite size scaling of the fourth order cumulant for $q=6$.
(a) $V_L$ as a function of temperature for different system sizes. Averages were taken over several samples ranging from $300$ to $400$ for small
system sizes down to $50$ and $20$ for $L=128$ and $L=256$. The orange line indicate the analytically predicted location of the minimum in the thermodynamic limit.
(b) Pseudo critical temperature $T_c^*$ {\it vs.} $1/L^2$. Error bars, estimated from the uncertainty when locating the minimum of $V_L$,
are shown only when larger than the symbol size.
}
\label{VL-q6}
\end{figure}

\noindent In Fig.\ref{VL-q6}b we show $T_c^*(L)$ vs. $1/L^2$ for $q=6$. The extrapolated value of $T_c^*(L)$ for $L\to\infty$,
$0.8078\pm0.0002$ agrees with the exact value $T_c=0.8076068...$ within an accuracy of the $0.025\%$.
%
%

Let us emphasize that, as it is well known, it's very difficult to get good measures
of cumulants with a single spin flip MC algorithm. In order to get reliable averages of
the cumulant minimum location,
one should guarantee a measurement time long enough to let the system
overcome the phase separating energy barrier back and forward several times.
Moreover, the characteristic activation time to overcome the barrier increases both with $q$ and $L$ (it increases exponentially with $L$).
For instance, simulation times of the order $10^7$ for each temperature are needed to obtain a good sampling for $q=6$ and $L=256$.

\begin{figure}
\centering
\includegraphics[scale=0.45]{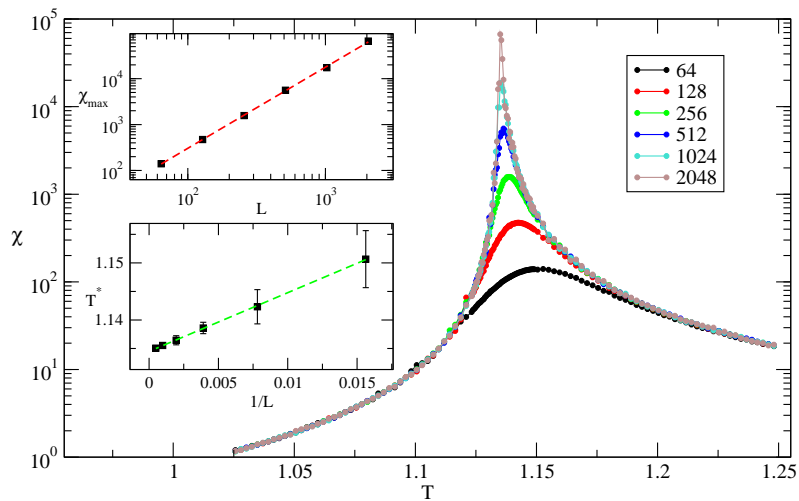}
\caption{(Color online) Finite size scaling of the susceptibility for $q=2$.
(main plot) $\chi$ as a function of temperature for different system linear sizes. Averages were taken over several samples ranging from $300$ for small
system sizes down to $50$ and $15$ for $L=1024$ and $L=2048$, respectively. We have used equally equilibration and measurement times of $2\times10^5 \mathit{MCS}$,
measuring quantities each $10 \mathit{MCS}$, thus totalizing averages over $6\times 10^6$ to $3\times 10^5$ as we increase the system size.
(upper inset) Maximum value of the susceptibility peak $\chi_{max}$ {\it vs.} $L$. Error bars, estimated from the uncertainty when evaluating the maximum,
are smaller than the symbol size.
(lower inset) Pseudo critical temperature $T^*(L)$ {\it vs.} $1/L$. Error bars, estimated from the uncertainty when locating the position of the maximum,
are shown only when larger than the symbol size.
}
\label{q2chi}
\end{figure}

In addition, we test our code for the $q=2$ (Ising) case. Fig.\ref{q2chi} shows the susceptibility of the order parameter calculated as
\begin{equation}
\chi  = \frac{N}{T} \left[\left<m^2\right> - \left<m\right>^2\right]
\end{equation}

\noindent The extrapolated value of the pseudo critical temperature $T^*(L)$ (defined as the location of the susceptibility maximum) for $L\to\infty$,
$1.1345\pm0.0001$, agrees with the exact value\footnote{ It should be remembered that $J_{Potts} = 2 J_{Ising}$ if we compare our hamiltonian
(\ref{Hamiltonian}) with the usual Ising hamiltonian, thus giving a $T_c(q=2)$ which is a half of the commonly appearing in Ising model works.}
$T_c(q=2)=1.1345926...$ within an accuracy of the $0.009\%$.
Even more, if we plot the maximum value of $\chi$ against the linear size $L$ it is expected to observe a finite size scaling of the form $\chi_{max} \sim L^{\gamma/\nu}$
\cite{Landau-Binder-2009}, where $\gamma$ and $\nu$ are the exactly known critical exponents for the 2D Ising model. We obtain such scaling with a combined exponent
$\gamma/\nu = 1.77\pm0.02$,  in a good agreement with the exact value $\gamma/\nu = \frac{7/4}{1} = 1.75$.


\section{Algorithm performance}
\label{performance}


The first step towards performance analysis is the kernel function calling breakdown.
In this case, it is done using CUDA profiling capabilities and some scripting to analyze a
2.9GB \code{cuda\_profile\_0.log} file produced after $12,6$ hours of computation.
The parameters used for this profiling are $q=9, N=2048\times 2048$, $T_{\mathit{min}} = 0.721200$,
$T_{\mathit{max}} = 0.721347$, $\delta T = 10^{-5}$, $t_{tran}=10^{5} \mathit{MCS}$,
$t_{max}=10^{4} \mathit{MCS}$ and $\delta t=500 \mathit{MCS}$.

The profile shows that there are approximately 32 millions of calls to \code{updateCUDA} and just a few
thousands to the other three kernels.
Since the individual gpu time consumptions of each kernel are comparable,
the only relevant kernel to analyze is \code{updateCUDA}.

To analyze the kernel \code{updateCUDA} we sweep $L$ in the range from $512$ to $32768$ in powers of two,
measuring the average execution time of the kernel and normalizing it to nanoseconds per spin flip.

We compare the three GPUs, using the same machine code (Compute Capability -- CC 1.3, generated by
NVCC 3.2)\footnote{Using CC 2.0 ISA does not bring any performance improvement.}, and the same video driver (driver version 260.19).
We also compare the GPUs performance with a CPU implementation.
For this version, we tried to keep the structure of the CUDA code, in order to compare the execution of the same physical protocol
on each architecture. We replaced the calls to CUDA kernels with loops running over all the spins in the same checkerboard scheme, we used
the same MWC random number generator.
We also added some optimizations to improve the CPU performance like creating a precomputed table of Boltzmann weights for the spinflip acceptance
for each simulated temperature, since the CPU have no mechanism for hiding memory latency and the impact of any floating-point unit (FPU)
computation is noticeable.
We run the CPU code against a Core 2 Duo architecture (E8400 -- Q1 2008) using GCC 4.4.5 with carefully chosen optimization
flags\footnote{Compiler options \code{-O3 -ffast-math -march=native -funroll-loops}.}.

We also vary $q$ in the set $\{6,8,9,12,15,24,48,96,192\}$. We don't find any significant variation of the performance with $q$,
except in the $q=2^k$ cases for the GTX 280, where the compiler obtains slight performance advantages using bitwise operators for modulus operation. The Fermi board has an improved modulus, rendering that difference imperceptible.

The profiling measurement is done in the GPU cases using CUDA profiling capabilities that gives very precise results,
avoiding any code instrumentation.
For the CPU version it is necessary to instrument the code with simple system calls to obtain the wall time.
In order to make the measurement independent of the temperature range covered, given that the transition temperature (and therefore
the flip acceptance rate) changes with $q$, we choose a deterministic write, i.e. we always write the spin value irrespective if the spin
changes its state respect of its previous state or not.
Writing the spin value only when it changes its state, brings a slight performance improvement around 2\% in the general case.

\begin{figure}[h!]
\centering
\includegraphics[scale=0.45]{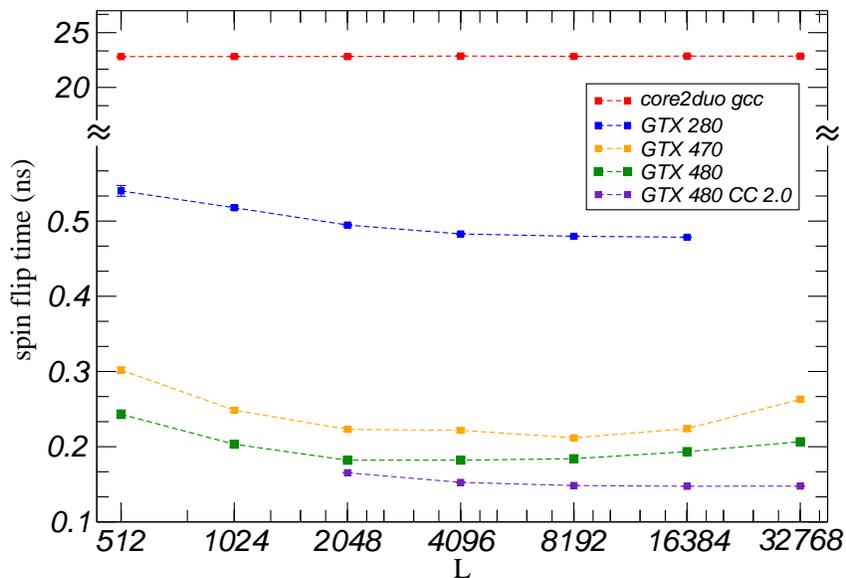}
\caption{Spin flip time in nanoseconds vs. lattice size running on an Intel Core 2 Duo E8400@3.0GHz CPU, and running on GTX 280, GTX 470 and GTX 480 NVIDIA GPUs. Averages are performed over $400$ runs for the GPUs and $60$ runs for the CPU. Error bars are smaller than symbol sizes when not showed.
}
\label{figure:tsfL}
\end{figure}

In figure~\ref{figure:tsfL} we can see that the curve corresponding to the CPU implementation is flat around\footnote{
It's worth mentioning that in order to compare this value with CPU implementations of the Ising model (e.g., 8ns in \cite{We2011}),
one should take into account that the Potts model update routine requires an extra random number to choose where to flip the spin.
In addition, using MWC doesn't provide the fastest execution times; other RNGs as LCG-32 give better times but not completely reliable
results \cite{We2011} due to their short period. For the sake of completeness, we report that eliminating one random number toss and
using LCG-32 instead of MWC we obtain a spin flip time of 14.5ns for our CPU implementation.
}
22.8ns, showing no dependence of the averaged spin flip time with system size.
For GPU cases, instead, we do have variations with respect to $L$.
The slowest card is the GTX 280, with spin flip times in the range [0.48ns, 0.54ns] which are
47x to 42x faster than those of the CPU code.
The GTX 470 has a variation between 0.21ns and 0.30ns, giving a speedup between 108x and 76x.
The fastest card is the GTX 480 with spin flip times in [0.18ns, 0.24ns] achieving a speedup from 126x to 95x.
There is also another curve corresponding to a specifically tuned version for the GTX 480 card\footnote{Each block is filling
the maximum 1024 threads, we also disable L1 cache for a (free) slight performance improvement: compiler options \code{-Xptxas -dlcm=cg -Xptxas -dlcm=cg}.} and CC 2.0, obtaining 155x (0.147ns) for the fastest case.
It is important to notice that even when using newer CPU architectures like Nehalem (X5550 -- Q1 2009) the spin flip time only
drops 2ns in the best case respect to the Core 2 Duo, and that Intel C++ Compiler (ICC) cannot do any better than that.

Nevertheless, it should be noted that better CPU implementations could be possible, since most appropriate implementations
for each architecture could be quite different from each other. For example, lower times can be attained for CPU using typewriter update
scheme instead of a checkerboard one.
For that reason, we hold the idea that a good measure to compare performances between GPU implementations is the ``time per spin flip'',
and the speedup respect to a CPU implementation is just additional illustrative information.

The variations for the GPU cards are due to two competing factors in the loop of the update kernel.
One is strictly decreasing with $L$ and is related to the amount of global memory movements per cell.
Since there is one RNG for each thread, the global memory for the RNG state is retrieved one time in the
beginning and stored in the end, therefore the larger the $L$, this single load/store global memory latency
is distributed into more cells.
The second factor is increasing in $L$ and is given by the inherent overhead incurred by a loop
(comparison and branching), that for $L=32768$ amounts to 4096 repetitions.

We also frame at $256\times 256$ and $1024\times 1024$, obtaining a 25\% of performance penalty for the former, and
a performance increase of 2\% in the later.
This gives us more evidence that the framing at $512\times 512$ is an appropriate trade-off between memory consumption
by the RNG and the speed of the code.

Although there are divergent branches inside the code, even for deterministic cell writes (the boolean ``or'' operator
semantics is shortcircuted), eliminating all divergent branches doing an arithmetic transformation does not bring any
performance improvement.
This shows the dominance of memory requests over the integer and floating point operations, and the ability of the hardware
scheduler in hiding the divergent branch performance penalty in between the memory operations.

\medskip

To our knowledge this is the first time the Potts model is implemented in GPUs, so there is no direct performance comparison.
There exist previous works that deal with similar problems and that report performance measurements.
Preis \textit{et. al}~\cite{PrViPaSc2009} implemented a 2D Ising model in GPUs, they reported a speedup of $60x$
upon their CPU implementation using a GTX 280. Their implementation has the disadvantage that the system size is limited by the \emph{maximum number of threads per block} allowed (enforcing $L\leq 1024$ on GT200 and $L\leq 2048$ on GF100). Later on, Block, Virnau and Preis~\cite{BlViPr2010} simulated the 2D Ising model using multi-spin coding techniques obtaining 0.126ns per spin flip in a GT200 architecture.
Weigel~\cite{We2010,We2011} has also considered the 2D Ising model, obtaining a better 0.076ns per spin flip~\cite{weigel10site} on the same architecture, which is improved to 0.034ns per spin flip on a Fermi (GF100) architecture. Moreover, this was obtained with a single-spin coded
implementation; however this gain is partially due to the use of a multi-hit technique updating up to $k=100$ times a set of cells while others
remain untouched.
Notwithstanding, Weigel obtains~\cite{We2011} 0.13ns per spin flip for the update without multi-hit and multi-spin, which is comparable with
the result of the multi-spin coded version in~\cite{BlViPr2010}.
Performance results on the 3D Ising model are also available~\cite{PrViPaSc2009, We2011}.
The Heisenberg spin glass model is simulated on a GPU in Ref.\cite{BePaPa2011}, and for this floating point vector spin, they achieve a 0.63ns
per spin flip update on a GF100 architecture.
%
Implementations of the Heisenberg model are also reported in \cite{We2011} with times per spin flip down to 0.18ns on a Fermi architecture,
representing impressive speedups (up to 1029x).
Recently, a GPU parallelization for the GF200 architecture was implemented in the Cellular Potts Model~\cite{TaSo2011}
with $\sim 80x$ speedup respect to serial implementations.


We also conduct end-to-end benchmarks of a small simulation
($q\!=\!9$, $L\!=\!1024$, \# of samples=3, $T_{min}\!=\!0.71$, $T_{max}\!=\!0.73$,
$\delta T\!=\!0.002$, $t_{tran}\!=\!2000$, $t_{max}\!=\!8000$, $\delta t\!=\!50$).
We obtain 193s for the GTX 280 and 8115s for the Intel Core 2 architecture, with a global speedup of 42x,
very similar to the speedup reported by the microbenchmarks.
The coincidence of microbenchmarks and end-to-end benchmarks results
reaffirms the fact that all the optimization efforts should go to the update kernel
\code{updateCUDA}.

\section{Metastability in the q-state Potts model}
\label{metastability}

Based on Binder's criterion described in Section~\ref{model} we analyze the existence of metastability for $q>4$ as the system size increases.
From Fig.\ref{fig3} we see that  for large enough values of  $q$  the energy branches attain
the transition temperature from both sides with a finite slope, even with a relatively poor temperature resolution. As $q$ decreases, a closer approach to $T_c$ is needed in order to distinguish whether a true singularity at $T_c$ is present or not, since the spinodal temperatures are expected to be located very close to~\cite{LoFeGrCa2009} $T_c$.

A power law divergence of the specific heat at $T_c$ would imply the following behavior
\begin{eqnarray}
    e_{T<T_c}  & = & e_o - A^- (1-T/T_c)^{1-\alpha_-} \label{divergence-at-Tc}\\
    e_{T>T_c}  & = & e_d - A^+ (1-T_c/T)^{1-\alpha_+}\label{divergence-at-Tc-2}
\end{eqnarray}
with $\alpha_-,\alpha_+ > 0$\\

On the other hand, if well defined metastable states occur, the energy
could be represented in terms of a specific heat diverging at pseudospinodal
temperatures $T_{sp}^+ , T_{sp}^-$
\begin{eqnarray}
    e_{T<T_c} & = & e_{sp}^- - A^- (1-T/T_{sp}^+)^{1-\alpha_-} \label{divergence-at-Tsp}\\
    e_{T>T_c} & = & e_{sp}^+ - A^+ (1-T_{sp}^-/T)^{1-\alpha_+}\label{divergence-at-Tsp-2}
\end{eqnarray}

If divergences for the specific heat occur at the pseudospinodals, we should
see exponents $\alpha_- = \alpha_+ \approx 0$ in Eqs.(\ref{divergence-at-Tc}) and (\ref{divergence-at-Tc-2}),  since
Eqs.(\ref{divergence-at-Tsp}) and (\ref{divergence-at-Tsp-2}) imply  finite slopes at $T_c$.\\

\begin{figure}
\label{pendsup}
\centering
\includegraphics[scale=0.45]{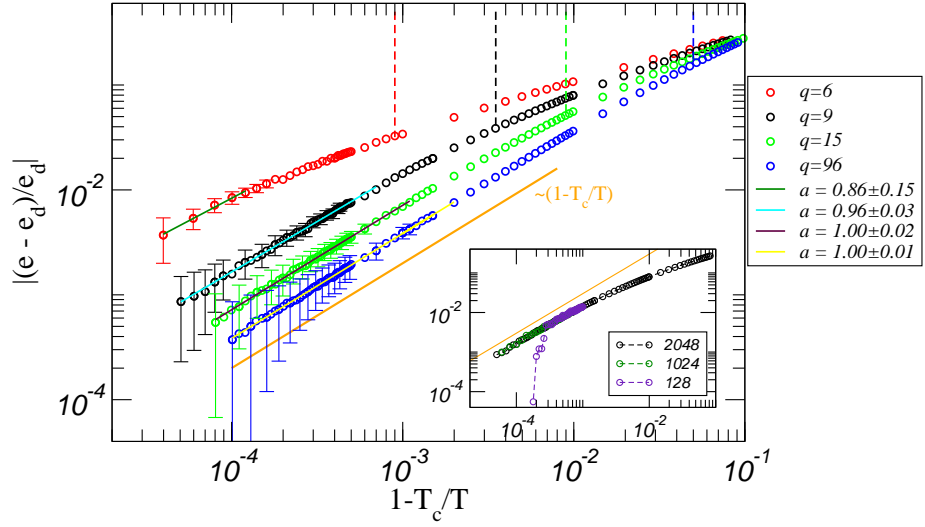}
\caption{(Color online) Log-log plot of energy differences versus temperatures $T>T_c$ for various $q$.
Data correspond to averages over $20$ samples of systems size $L=2048$,
equilibration times ranging from $~5 \times 10^4[MCS]$ to $~2\times 10^5[MCS]$ and measurement times  of $~5 \times 10^4[MCS]$, with sampling every $100[MCS]$. Error bars were estimated considering a $90\%$ confidence interval (only some representative error bars are shown for clarity).
Full color lines are power-law fits of the form $|(e-e_d)/e_d| = A (1-T_c/T)^a$ (resulting exponents $a$ are showed in the labels).
Dashed vertical lines of different colors indicate correspond to $T=T_c+ \Delta T(q)$, with $\Delta T =T_c-T_{sp}^-$  and $T_{sp}^-$ from Eq.(\ref{Tsp-STD}). The inset shows $q=9$ curves for
different system sizes, the full orange curve indicates the slope 1.}
\end{figure}

\begin{figure}
\label{pendinf}
\centering
\includegraphics[scale=0.45]{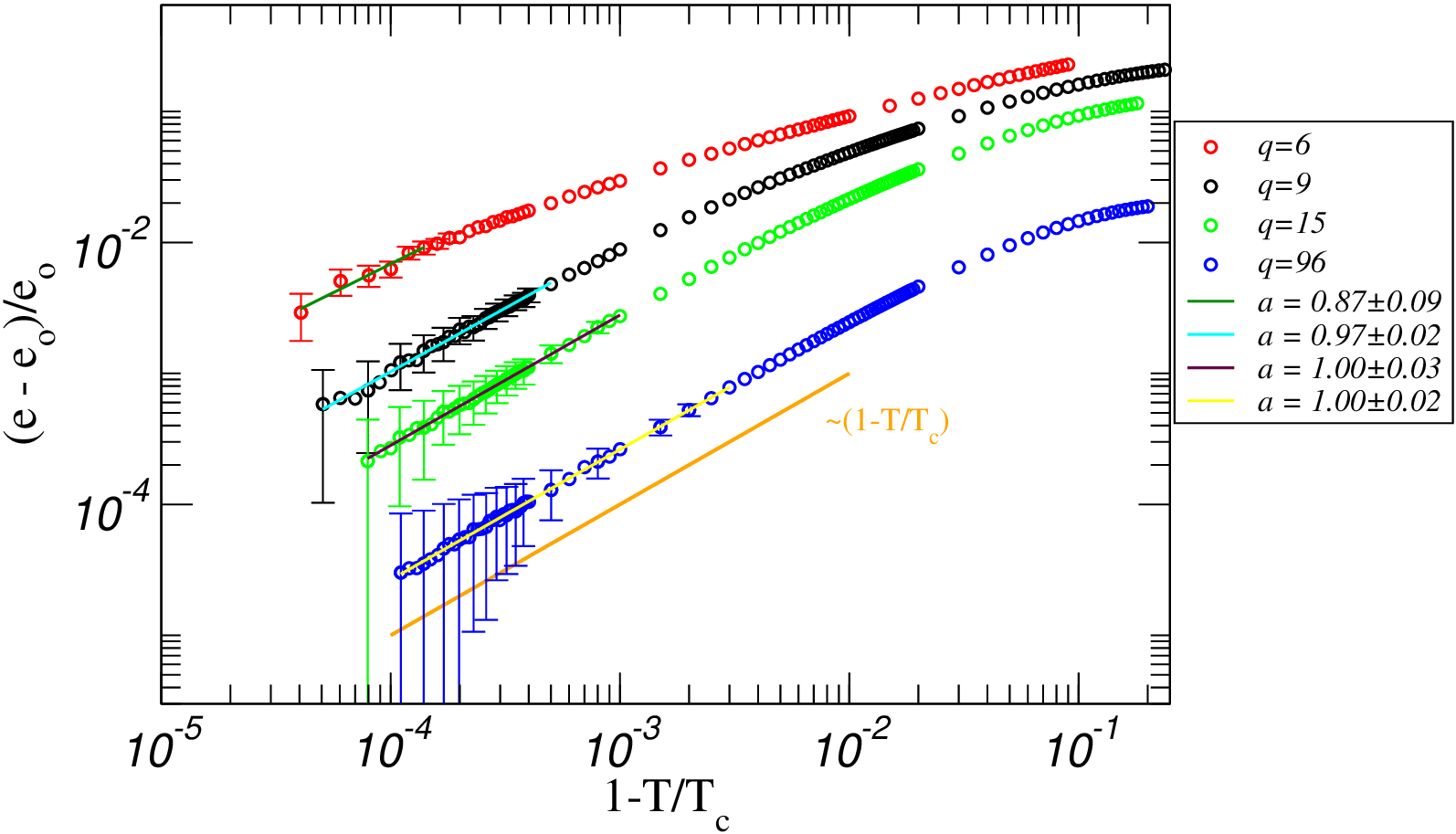}
\caption{(Color online) Log-log plot of energy differences versus temperatures $T<T_c$ for various $q$. Data correspond to averages over $20$ samples of systems size $L=2048$,
equilibration times ranging from $~5 \times 10^4[MCS]$ to $~2\times 10^5[MCS]$ and measurement times  of $~5 \times 10^4[MCS]$, with sampling every $100[MCS]$. Error bars were estimated considering a $90\%$ confidence interval (only some representative error bars are shown for clarity).
Full color lines are power-law fits of the form $(e-e_o)/e_o = A (1-T/T_c)^a$ (resulting exponents $a$ are showed in the labels).
}
\end{figure}

We measure equilibrium curves for $e_{T<T_c}$ ($e_{T>T_c}$) starting from a
ordered (disordered) initial state and performing a cooling (heating) procedure approaching $T_c$, as described in section \ref{algorithm}.
The results are presented in Fig.\ref{pendsup} and~\ref{pendinf}.
In both figures a crossover
of the curve's slope as we approach $T_c$ can be observed for all values of $q$.
Close enough to $T_c$, the curves for $q=9,15,96$ show exponents which are indistinguishable from $1$, consistently with the existence of metastability and divergences at spinodal temperatures different from $T_c$, at least for $q \geq 9$.

As pointed out by Binder \cite{Bi1981}, to observe the crossover (if it exists at all) a temperature resolution at least $\Delta T =T_c-T_{sp}^-$ for the high energy branch (or $\Delta T =T_{sp}^+ - T_c$ for the low energy branch) is needed, where $\Delta T \equiv |T-T_c|$. A numerical estimation of the lower spinodal temperature predicted by Short Time Dynamics \cite{LoFeGrCa2009} is given by

\begin{equation}\label{Tsp-STD}
 \frac{T_c-T_{sp}^-}{T_c}\simeq 0.0007 \left(\ln(1+q-4)\right)^{2.81}.
\end{equation}

\noindent The vertical dashed lines in Fig.\ref{pendsup} correspond to $T=T_c + \Delta T(q)$, as predicted from Eq.(\ref{Tsp-STD}) according to the previous criterion. The coincidence with the crossover points for all values of $q$ shows a complete agreement between the present results and those from Short Time Dynamics calculations. To attain the desired temperature resolution the system size has to be large enough, since finite size rounding errors are expected to decay as $1/L$ \cite{Bi1981,Ja1993}. This is illustrated  in the inset of Fig.\ref{pendsup} for the particular case $q=9$, where a strong finite size effect is observed for $L=128$.  A rough estimation  of the minimum size required to reduce the error $L \approx 1/\Delta T$ predicts $L=400$. We see that this finite size effect is suppressed for sizes $L \geq 1000$. Moreover,  further increase of  the system size does not change the behavior of the curves close to $T_c$.

We have no  estimations for $T_{sp}^+$ for arbitrary values of $q$, but a close look to the
curves in Fig.\ref{fig3} suggest that $T_{sp}^+$ is closer to $T_c$ than $T_{sp}^-$ is.
This is consistent with the behavior observed in Fig.\ref{pendinf},
where crossovers occur closer to $T_c$ than in Fig.\ref{pendsup}.

Our results  for $q=6$ are not conclusive. For instance, in the high energy branch we observe the previously discussed crossover, but the slope  changes from $0.6$ to $0.8$. Such variation is of the same order of the fitting error below the crossover. This is because statistical fluctuations in the energy become very important at the required temperature resolution level ($\Delta T/T_c \leq 10^{-4}$), as can be seen in Fig.\ref{pendsup}. Hence, to obtain a clear answer a very large sample size (one can roughly estimate $\sim 2000$) and probably a larger system size  is needed. In fact, we performed simulations with a sample size $50$ (for $L=2048$), without any improvement in the results.
We even simulate systems of $L=8192$ with a sample size on the order of $10$ with no appreciable change.

The situation is more difficult for the low energy branch, where no clear evidence of crossover is observed (see Fig.\ref{pendinf}). However, one could expect  the existence of an upper spinodal temperature $T_{sp}^+$  located closer to $T_c$ than the lower one $T_{sp}^-$ and therefore a higher temperature resolution (together with  larger system and sampling sizes) would be needed to elucidate whether there is metastability or not.

\section{Discussion}
\label{discussion}

We implemented a CUDA-based parallel Monte Carlo algorithm to simulate the Statistical Mechanics of the q-state Potts model.
The code allows a speedup (compared with an optimized serial code running on a CPU) from 42x in the GTX 280 card up to 155x in a GTX 480, with an average time per spin flip of 0.54ns down to 0.147ns respectively.
Those times are of the same order of previous implementations in the simpler case of the Ising model, without the usage of sophisticated programming techniques, such as multi spin coding.
Besides the speedup, the present algorithm allows the simulation of very large systems in very short times, namely $\sim10^9$ spins with an average time per $\mathit{MCS}$ of 0.15s.
Such performance is almost independent of the value of $q$.
The key factors to achieve those numbers is the per-thread independent RNG that is fast and takes only a few registers, the framing scheme that increases the amount of computation done by each thread and at the same time it bounds the number of independent RNG needed, and finally the cell-packing mapping that orders the memory access.

The possibility of performing high speed simulations at large enough system sizes  allowed us to study the metastability problem in the two dimensional system based on Binder's criterion, namely, on the existence or not of specific heat singularities at spinodal temperatures different from the transition one (but very close to). Our results provide a positive numerical evidence about the existence of metastability on very large systems, at least for $q\geq 9$.

Even when our results for $q=6$ suggest the same behavior as for larger values of $q$, they could also be consistent with the absence of metastability. Hence, one cannot exclude the existence of a second critical value $4<q^*\leq 9$ such that metastability disappears when $4<q<q^*$.

Although the present implementation was done for a two dimensional system with nearest neighbors interactions (checkerboard update scheme), its generalization to three dimensional systems and/or longer ranged interactions is feasible, but some features should be adjusted.
For the generalization to the 3D case, the checkerboard scheme defining two independent sub-networks persists, however the cell-packing scheme should be updated conveniently.
For  the 2D case with first and second neighbors interactions, there are nine
independent sub-networks to update instead of two.
The combination of both generalizations is direct.

The present implementation is based on the simplest single-spin flip algorithm namely, Metropolis. Its extension to more sophisticated single spin flip algorithms (See for example Refs.\cite{LoQiScJi2004}, \cite{SuTo2010}) is also straightforward and represent an interesting prospective in the field.
 In particular, temperature reweighting~\cite{FeSw1989} or other histogram based techniques (see for example~\cite{Landau-Binder-2009})
can be implemented by keeping track of the energy changes at each spin flip for each step,
instead of making the calculation of the energy over the whole system at each step. This kind of tracking could be done without loose of performance
by implementing a paralell acumulation of local energy changes \emph{on-the-fly} taking advantage of the GPU's hierarchic memory scheme.

Besides its theoretical interest, the large-$q$ Potts model (or minor variations of it) is widely
used for simulating the dynamics of a
large variety of systems, such as soap bubbles and foam~\cite{GlWe1992,SaGl2006}, grain growth~\cite{WeGl1992,ThAlGr2006}, gene segregation~\cite{KoAvHaOsNe2010}, biological cells~\cite{GrGl1992}, tumor migration~\cite{TuSh2002}, image segmentation~\cite{Be2010}, neural networks~\cite{Kr2008} and social demographics behavior~\cite{Sc2005,TrBr2009}. The present implementation of the Potts model on GPUs, or easy modifications of it,
would result helpful for some of the above cited applications. The possibility of simulating
bigger systems and having results faster than usual should be welcomed
in the statistical physics community. Our CUDA code is available for download and use under GNU GPL 3.0 at our Group webpage~\cite{potts3site}.

\bigskip

\paragraph{Acknowledgments}
We  thank C. Bederi\'an for very useful suggestions.
We would also like to thank to A. Kolton and C. S\'anchez for kindly giving access to
a GTX 470 and a GTX 480 respectively, for benchmarks and simulations.
Fruitful discussions and suggestions from O. Reula and M. Bellone are also acknowledged. This work was partially supported by grants from
FONCYT/ANPCYT (Argentina), CONICET (Argentina), SeCyT, Universidad Nacional de C\'ordoba
(Argentina) and NVIDIA professor partnership program.

\bibliographystyle{model1a-num-names}

\begin{thebibliography}{58}
\expandafter\ifx\csname natexlab\endcsname\relax\def\natexlab#1{#1}\fi
\providecommand{\bibinfo}[2]{#2}
\ifx\xfnm\relax \def\xfnm[#1]{\unskip,\space#1}\fi
\bibitem[{Fischer and Hertz(1993)}]{Fisher-Hertz-book1993}
\bibinfo{author}{K.~H. Fischer}, \bibinfo{author}{J.~A. Hertz},
  \bibinfo{title}{Spin Glasses}, \bibinfo{year}{1993}.
\bibitem[{{Kob}(2003)}]{Ko2003}
\bibinfo{author}{W.~{Kob}}, \bibinfo{journal}{in Lecture notes for ``Slow
  relaxations and nonequilibrium dynamics in condensed matter'', Les Houches
  Session LXXVII}  (\bibinfo{year}{2003}) \bibinfo{pages}{199}.
\bibitem[{Binder and Kob(2005)}]{Binder-Kob-book2005}
\bibinfo{author}{K.~Binder}, \bibinfo{author}{W.~Kob}, \bibinfo{title}{Glassy
  Materials and Disordered Solids: An Introduction to Their Statistical
  Mechanics}, \bibinfo{publisher}{World Scientific Publishing - Singapore},
  \bibinfo{year}{2005}.
\bibitem[{{Cugliandolo}(2010)}]{Cu2010}
\bibinfo{author}{L.~F. {Cugliandolo}}, \bibinfo{journal}{Physica A Statistical
  Mechanics and its Applications} \bibinfo{volume}{389} (\bibinfo{year}{2010})
  \bibinfo{pages}{4360--4373}.
\bibitem[{Preis et~al.(2009)Preis, Virnau, Paul, and Schneider}]{PrViPaSc2009}
\bibinfo{author}{T.~Preis}, \bibinfo{author}{P.~Virnau},
  \bibinfo{author}{W.~Paul}, \bibinfo{author}{J.~J. Schneider},
  \bibinfo{journal}{Journal of Computational Physics} \bibinfo{volume}{228}
  (\bibinfo{year}{2009}) \bibinfo{pages}{4468 -- 4477}.
\bibitem[{Block et~al.(2010)Block, Virnau, and Preis}]{BlViPr2010}
\bibinfo{author}{B.~Block}, \bibinfo{author}{P.~Virnau},
  \bibinfo{author}{T.~Preis}, \bibinfo{journal}{Computer Physics
  Communications} \bibinfo{volume}{181} (\bibinfo{year}{2010})
  \bibinfo{pages}{1549 -- 1556}.
\bibitem[{Bernaschi et~al.(2011)Bernaschi, Parisi, and Parisi}]{BePaPa2011}
\bibinfo{author}{M.~Bernaschi}, \bibinfo{author}{G.~Parisi},
  \bibinfo{author}{L.~Parisi}  \bibinfo{journal}{Computer Physics
  Communications} \bibinfo{volume}{182} (\bibinfo{year}{2011})
  \bibinfo{pages}{1265 -- 1271}.
\bibitem[{Hawick et~al.(2010)Hawick, Leist, and Playne}]{HaLePl2010}
\bibinfo{author}{K.~Hawick}, \bibinfo{author}{A.~Leist},
  \bibinfo{author}{D.~Playne}, \bibinfo{journal}{International Journal of
  Parallel Programming}  (\bibinfo{year}{2010}) \bibinfo{pages}{1--19}.
\bibitem[{Weigel(2010)}]{We2010}
\bibinfo{author}{M.~Weigel}, \bibinfo{journal}{Computer Physics Communications}
  \bibinfo{volume}{182} (\bibinfo{year}{2011})
  \bibinfo{pages}{1833 -- 1836}.
\bibitem[{Weigel(2011)}]{We2011}
\bibinfo{author}{M.~Weigel},
  \bibinfo{note}{ArXiv:1101.1427},
  \bibinfo{year}(2011).
\bibitem[{Herrmann et~al.(2010)Herrmann, Silberholz, Bellone, Guerberoff, and
  Tiglio}]{HeSiBeGuTi2010}
\bibinfo{author}{F.~Herrmann}, \bibinfo{author}{J.~Silberholz},
  \bibinfo{author}{M.~Bellone}, \bibinfo{author}{G.~Guerberoff},
  \bibinfo{author}{M.~Tiglio}, \bibinfo{journal}{Classical and Quantum Gravity}
  \bibinfo{volume}{27} (\bibinfo{year}{2010}) \bibinfo{pages}{032001}.
\bibitem[{Tickner(2010)}]{Ti2010}
\bibinfo{author}{J.~Tickner}, \bibinfo{journal}{Computer Physics
  Communications} \bibinfo{volume}{181} (\bibinfo{year}{2010})
  \bibinfo{pages}{1821 -- 1832}.
\bibitem[{Clark et~al.(2010)Clark, Babich, Barros, Brower, and
  Rebbi}]{ClBaBaBrRe2010}
\bibinfo{author}{M.~Clark}, \bibinfo{author}{R.~Babich},
  \bibinfo{author}{K.~Barros}, \bibinfo{author}{R.~Brower},
  \bibinfo{author}{C.~Rebbi}, \bibinfo{journal}{Computer Physics
  Communications} \bibinfo{volume}{181} (\bibinfo{year}{2010})
  \bibinfo{pages}{1517 -- 1528}.
\bibitem[{Ryoo et~al.(2008)Ryoo, Rodrigues, Baghsorkhi, Stone, Kirk, and mei
  W.~Hwu}]{Ryoo08}
\bibinfo{author}{S.~Ryoo}, \bibinfo{author}{C.~I. Rodrigues},
  \bibinfo{author}{S.~S. Baghsorkhi}, \bibinfo{author}{S.~S. Stone},
  \bibinfo{author}{D.~B. Kirk}, \bibinfo{author}{W.~mei W.~Hwu}, in:
  \bibinfo{booktitle}{Proceedings of the ACM SIGPLAN Symposium on Principles
  and Practice of Parallel Programming (23th PPOPP'2008)},
  \bibinfo{publisher}{ACM SIGPLAN}, \bibinfo{year}{2008}, pp.
  \bibinfo{pages}{73--82}.
\bibitem[{Wu(1982)}]{Wu1982}
\bibinfo{author}{F.~Y. Wu}, \bibinfo{journal}{Rev. Mod. Phys.}
  \bibinfo{volume}{54} (\bibinfo{year}{1982}) \bibinfo{pages}{235}.
\bibitem[{Binder(1981)}]{Bi1981}
\bibinfo{author}{K.~Binder}, \bibinfo{journal}{J. Stat. Phys.}
  \bibinfo{volume}{24} (\bibinfo{year}{1981}) \bibinfo{pages}{69}.
\bibitem[{Meunier and Morel(2000)}]{MeMo2000}
\bibinfo{author}{J.~Meunier}, \bibinfo{author}{A.~Morel},
  \bibinfo{journal}{Eur. Phys. J. B} \bibinfo{volume}{13}
  (\bibinfo{year}{2000}) \bibinfo{pages}{341}.
\bibitem[{Petri et~al.(2008)Petri, {Iba\~{n}ez de Berganza}, and
  Loreto}]{PeIbLo2008}
\bibinfo{author}{A.~Petri}, \bibinfo{author}{M.~{Iba\~{n}ez de Berganza}},
  \bibinfo{author}{V.~Loreto}, \bibinfo{journal}{Philosophical Magazine}
  \bibinfo{volume}{88} (\bibinfo{year}{2008}) \bibinfo{pages}{3931 -- 3938}.
\bibitem[{Bazavov et~al.(2008)Bazavov, Berg, and Dubey}]{BaBeDu2008}
\bibinfo{author}{A.~Bazavov}, \bibinfo{author}{B.~A. Berg},
  \bibinfo{author}{S.~Dubey}, \bibinfo{journal}{Nuclear Physics B}
  \bibinfo{volume}{802} (\bibinfo{year}{2008}) \bibinfo{pages}{421 -- 434}.
\bibitem[{Loscar et~al.(2009)Loscar, Ferrero, Grigera, and
  Cannas}]{LoFeGrCa2009}
\bibinfo{author}{E.~Loscar}, \bibinfo{author}{E.~Ferrero},
  \bibinfo{author}{T.~Grigera}, \bibinfo{author}{S.~Cannas},
  \bibinfo{journal}{Jour. Chem. Phys.} \bibinfo{volume}{131}
  (\bibinfo{year}{2009}) \bibinfo{pages}{024120}.
\bibitem[{Vinals and Grant(1987)}]{ViGr1987}
\bibinfo{author}{J.~Vinals}, \bibinfo{author}{M.~Grant},
  \bibinfo{journal}{Phys. Rev. B} \bibinfo{volume}{36} (\bibinfo{year}{1987})
  \bibinfo{pages}{7036}.
\bibitem[{Grest et~al.(1988)Grest, Anderson, and Srolovitz}]{GrAnSr1988}
\bibinfo{author}{G.~S. Grest}, \bibinfo{author}{M.~P. Anderson},
  \bibinfo{author}{D.~J. Srolovitz}, \bibinfo{journal}{Phys. Rev. B}
  \bibinfo{volume}{38} (\bibinfo{year}{1988}) \bibinfo{pages}{4752--4760}.
\bibitem[{Sire and Majumdar(1995)}]{SiMa1995b}
\bibinfo{author}{C.~Sire}, \bibinfo{author}{S.~Majumdar},
  \bibinfo{journal}{Phys. Rev. E} \bibinfo{volume}{52} (\bibinfo{year}{1995})
  \bibinfo{pages}{244}.
\bibitem[{Ferrero and Cannas(2007)}]{FeCa2007}
\bibinfo{author}{E.~Ferrero}, \bibinfo{author}{S.~Cannas},
  \bibinfo{journal}{Phys. Rev. E} \bibinfo{volume}{76} (\bibinfo{year}{2007})
  \bibinfo{pages}{031108}.
\bibitem[{Loureiro et~al.(2010)Loureiro, Arenzon, Cugliandolo, and
  Sicilia}]{LoArCuSi2010}
\bibinfo{author}{M.~P.~O. Loureiro}, \bibinfo{author}{J.~J. Arenzon},
  \bibinfo{author}{L.~F. Cugliandolo}, \bibinfo{author}{A.~Sicilia},
  \bibinfo{journal}{Phys. Rev. E} \bibinfo{volume}{81} (\bibinfo{year}{2010})
  \bibinfo{pages}{021129}.
\bibitem[{Rutkevich(2002)}]{Ru2002}
\bibinfo{author}{S.~Rutkevich}, \bibinfo{journal}{Int. Jour. Mod. Phys.}
  \bibinfo{volume}{13} (\bibinfo{year}{2002}) \bibinfo{pages}{495}.
\bibitem[{Bauer et~al.(2010)Bauer, Gull, Trebst, Troyer, and
  Huse}]{BaGuTrTrHu2010}
\bibinfo{author}{B.~Bauer}, \bibinfo{author}{E.~Gull},
  \bibinfo{author}{S.~Trebst}, \bibinfo{author}{M.~Troyer},
  \bibinfo{author}{D.~A. Huse}, \bibinfo{journal}{Journal of Statistical
  Mechanics: Theory and Experiment} \bibinfo{volume}{2010}
  (\bibinfo{year}{2010}) \bibinfo{pages}{P01020}.
\bibitem[{Baxter(1973)}]{Ba1973}
\bibinfo{author}{R.~J. Baxter}, \bibinfo{journal}{J. Phys. C}
  \bibinfo{volume}{6} (\bibinfo{year}{1973}) \bibinfo{pages}{L:445}.
\bibitem[{Kihara et~al.(1954)Kihara, Midzuno, and Shizume}]{KiMiSh1954}
\bibinfo{author}{T.~Kihara}, \bibinfo{author}{Y.~Midzuno},
  \bibinfo{author}{T.~Shizume}, \bibinfo{journal}{J. Phys. Soc. Japan}
  \bibinfo{volume}{9} (\bibinfo{year}{1954}) \bibinfo{pages}{681}.
\bibitem[{Baxter(1982)}]{Ba1982}
\bibinfo{author}{R.~J. Baxter}, \bibinfo{journal}{J. Phys. A}
  \bibinfo{volume}{15} (\bibinfo{year}{1982}) \bibinfo{pages}{3329}.
\bibitem[{Velytsky et~al.(2003)Velytsky, Berg, and Heller}]{VeBeHe2003}
\bibinfo{author}{A.~Velytsky}, \bibinfo{author}{B.~A. Berg},
  \bibinfo{author}{U.~M. Heller}, \bibinfo{journal}{Nuclear Physics B -
  Proceedings Supplements} \bibinfo{volume}{119} (\bibinfo{year}{2003})
  \bibinfo{pages}{861 -- 863}.
\bibitem[{{Iba\~{n}ez de Berganza} et~al.(2007){Iba\~{n}ez de Berganza},
  Loreto, and Petri}]{IbLoPe2007}
\bibinfo{author}{M.~{Iba\~{n}ez de Berganza}}, \bibinfo{author}{V.~Loreto},
  \bibinfo{author}{A.~Petri}  (\bibinfo{year}{2007}).
  \bibinfo{note}{ArXiv:0706.3534}.
\bibitem[{Meyer-Ortmanns(1996)}]{Me1996}
\bibinfo{author}{H.~Meyer-Ortmanns}, \bibinfo{journal}{Rev. Mod. Phys.}
  \bibinfo{volume}{68} (\bibinfo{year}{1996}) \bibinfo{pages}{473--598}.
\bibitem[{Karsch and Stickan(2000)}]{KaSt2000}
\bibinfo{author}{F.~Karsch}, \bibinfo{author}{S.~Stickan},
  \bibinfo{journal}{Physics Letters B} \bibinfo{volume}{488}
  (\bibinfo{year}{2000}) \bibinfo{pages}{319 -- 325}.
\bibitem[{Bonati and D'Elia(2010)}]{BoEl2010}
\bibinfo{author}{C.~Bonati}, \bibinfo{author}{M.~D'Elia}
  (\bibinfo{year}{2010}). \bibinfo{note}{ArXiv:1010.3639}.
\bibitem[{Press et~al.(1992)}]{pres92nrinc}
\bibinfo{author}{W.~Press}, et~al., \bibinfo{title}{Numerical Recipes in {C}
  (Second Edition)}, \bibinfo{year}{1992}.
\bibitem[{Kernigham and Ritchie(1978)}]{KR78}
\bibinfo{author}{B.~W. Kernigham}, \bibinfo{author}{D.~M. Ritchie},
  \bibinfo{title}{The {C} Programming Language},
  \bibinfo{publisher}{PRENTICE-HALL, INC.}, \bibinfo{address}{Englewood Cliffs,
  NJ}, \bibinfo{year}{1978}.
\bibitem[{Kirk and mei W.~Hwu(2010)}]{Kirk10}
\bibinfo{author}{D.~B. Kirk}, \bibinfo{author}{W.~mei W.~Hwu},
  \bibinfo{title}{Programming massively parallel processors: a hands-on
  approach}, \bibinfo{publisher}{Morgan Kaufmann Publishers},
  \bibinfo{year}{2010}.
\bibitem[{Marsaglia(2003)}]{Marsaglia:2003:RNG}
\bibinfo{author}{G.~Marsaglia}, \bibinfo{journal}{Journal of Modern Applied
  Statistical Methods} \bibinfo{volume}{2} (\bibinfo{year}{2003})
  \bibinfo{pages}{2--13}.
\bibitem[{Erik~Alerstam(2009)}]{cudamcml_rng}
\bibinfo{author}{S.~A.-E. Erik~Alerstam, Tomas~Svensson},
  \bibinfo{title}{CUDAMCML User manual and implementation notes},
  \bibinfo{year}{2009}.
\bibitem[{Harris(2007)}]{harris07cuda}
\bibinfo{author}{M.~Harris}, \bibinfo{title}{Optimizing cuda},
  \bibinfo{howpublished}{SC07 Tutorial}, \bibinfo{year}{2007}.
\bibitem[{Merali(2010)}]{merali10error}
\bibinfo{author}{Z.~Merali}, \bibinfo{journal}{Nature} \bibinfo{volume}{467}
  (\bibinfo{year}{2010}) \bibinfo{pages}{775--777}.
\bibitem[{Barnes(2010)}]{barnes10publish}
\bibinfo{author}{N.~Barnes}, \bibinfo{journal}{Nature} \bibinfo{volume}{467}
  (\bibinfo{year}{2010}) \bibinfo{pages}{753--753}.
\bibitem[{pot(2010)}]{potts3site}
\bibinfo{howpublished}{\url{http://www.famaf.unc.edu.ar/grupos/GPGPU/Potts/CUD%
APotts.html}}, \bibinfo{year}{2010}. \bibinfo{note}{Q-State Potts model for
  CUDA site}.
\bibitem[{Janke(1993)}]{Ja1993}
\bibinfo{author}{W.~Janke}, \bibinfo{journal}{Phys. Rev. B}
  \bibinfo{volume}{47} (\bibinfo{year}{1993}) \bibinfo{pages}{14757--14770}.
\bibitem[{Binder(1997)}]{Bi1997}
\bibinfo{author}{K.~Binder}, \bibinfo{journal}{Rep. Prog. Phys.}
  \bibinfo{volume}{60} (\bibinfo{year}{1997}) \bibinfo{pages}{487}.
\bibitem[{Landau and Binder (2009)}]{Landau-Binder-2009}
\bibinfo{author}{D.~P. Landau}, \bibinfo{author}{K.~Binder},
  \bibinfo{title}{A Guide to Monte Carlo Simulations in Statistical Physics}
  (\bibinfo{edition}{third edition}), \bibinfo{year}{2009}.
\bibitem[{wei(2010)}]{weigel10site}
\bibinfo{howpublished}{\url{http://www.cond-mat.physik.uni-mainz.de/~weigel/re%
search/gpu-computing}}, \bibinfo{year}{2010}. \bibinfo{note}{Simulating spin
  models on GPU site}.
\bibitem[{Tapia and D'Souza(2011)}]{TaSo2011}
\bibinfo{author}{J.~J. Tapia}, \bibinfo{author}{R.~M. D'Souza},
  \bibinfo{journal}{Computer Physics Communications} \bibinfo{volume}{182}
  (\bibinfo{year}{2011}) \bibinfo{pages}{857 -- 865}.
\bibitem{LoQiScJi2004} D. Loison, C. L. Qin, K. D. Schotte and X. F. Jin, Eur. Phys. J. B {\bf 41}, 395–412 (2004).
\bibitem{SuTo2010}  H. Suwa and S. Todo, Phys. Rev. Lett. {\bf 105}, 120603 (2010).
\bibitem[{Ferrenberg and Swendsen(1989)}]{FeSw1989}
\bibinfo{author}{A.~M. Ferrenberg}, \bibinfo{author}{R.~H. Swendsen},
  \bibinfo{journal}{Phys. Rev. Lett.} \bibinfo{volume}{63}
  (\bibinfo{year}{1989}) \bibinfo{pages}{1658--1658}.
\bibitem[{Glazier and Weaire(1992)}]{GlWe1992}
\bibinfo{author}{J.~A. Glazier}, \bibinfo{author}{D.~Weaire},
  \bibinfo{journal}{J. Phys: Condens. Matter} \bibinfo{volume}{4}
  (\bibinfo{year}{1992}) \bibinfo{pages}{1867--1894}.
\bibitem[{Sanyal and Glazier(2006)}]{SaGl2006}
\bibinfo{author}{S.~Sanyal}, \bibinfo{author}{J.~A. Glazier},
  \bibinfo{journal}{Journal of Statistical Mechanics: Theory and Experiment}
  \bibinfo{volume}{2006} (\bibinfo{year}{2006}) \bibinfo{pages}{P10008}.
\bibitem[{Weaire and Glazier(1992)}]{WeGl1992}
\bibinfo{author}{D.~Weaire}, \bibinfo{author}{J.~A. Glazier},
  \bibinfo{journal}{Materials Science Forum} \bibinfo{volume}{94-96}
  (\bibinfo{year}{1992}) \bibinfo{pages}{27--38}.
\bibitem[{Thomas et~al.(2006)Thomas, de~Almeida, and Graner}]{ThAlGr2006}
\bibinfo{author}{G.~L. Thomas}, \bibinfo{author}{R.~M.~C. de~Almeida},
  \bibinfo{author}{F.~Graner}, \bibinfo{journal}{Phys. rev. E}
  \bibinfo{volume}{74} (\bibinfo{year}{2006}) \bibinfo{pages}{021407}.
\bibitem[{Korolev et~al.(2010)Korolev, Avlund, Hallatschek, and
  Nelson}]{KoAvHaOsNe2010}
\bibinfo{author}{K.~S. Korolev}, \bibinfo{author}{M.~Avlund},
  \bibinfo{author}{O.~Hallatschek}, \bibinfo{author}{D.~R. Nelson},
  \bibinfo{journal}{Rev. Mod. Phys.} \bibinfo{volume}{82}
  (\bibinfo{year}{2010}) \bibinfo{pages}{1691--1718}.
\bibitem[{Graner and Glazier(1992)}]{GrGl1992}
\bibinfo{author}{F.~Graner}, \bibinfo{author}{J.~A. Glazier},
  \bibinfo{journal}{Phys. Rev. Lett.} \bibinfo{volume}{69}
  (\bibinfo{year}{1992}) \bibinfo{pages}{2013--2016}.
\bibitem[{Turner and Sherratt(2002)}]{TuSh2002}
\bibinfo{author}{S.~Turner}, \bibinfo{author}{J.~A. Sherratt},
  \bibinfo{journal}{Journal of Theoretical Biology} \bibinfo{volume}{216}
  (\bibinfo{year}{2002}) \bibinfo{pages}{85 -- 100}.
\bibitem[{Bentrem(2010)}]{Be2010}
\bibinfo{author}{F.~Bentrem}, \bibinfo{journal}{Central European Journal of
  Physics} \bibinfo{volume}{8} (\bibinfo{year}{2010})
  \bibinfo{pages}{689--698}.
\bibitem[{Kryzhanovsky(2008)}]{Kr2008}
\bibinfo{author}{V.~Kryzhanovsky}, \bibinfo{journal}{in Artificial Neural
  Networks - ICANN 2008} \bibinfo{volume}{5164} (\bibinfo{year}{2008})
  \bibinfo{pages}{72--80}.
\bibitem[{Schulze(2005)}]{Sc2005}
\bibinfo{author}{C.~Schulze}, \bibinfo{journal}{Central European Journal of
  Physics} \bibinfo{volume}{16} (\bibinfo{year}{2005})
  \bibinfo{pages}{351--355}.
\bibitem[{Traag and Bruggeman(2009)}]{TrBr2009}
\bibinfo{author}{V.~A. Traag}, \bibinfo{author}{J.~Bruggeman},
  \bibinfo{journal}{Phys. Rev. E} \bibinfo{volume}{80} (\bibinfo{year}{2009})
  \bibinfo{pages}{036115}.

\end{thebibliography}

\end{document}